\documentclass[12pt]{article}

\usepackage{amsmath}
\usepackage{graphicx}
\usepackage{hyperref}
\newcommand{\Figref}[1]{Fig.~\ref{#1}}
\newcommand{\Figrefs}[1]{Figs.~\ref{#1}}
\newcommand{\Figrefstart}[1]{Figure~\ref{#1}}
\newcommand{\dd}{{\,\rm d}}

\newcommand{\Tabref}[1]{Table~\ref{#1}}

\usepackage{scicite}
\usepackage{epstopdf}
\usepackage{times}

\newenvironment{sciabstract}{%
\begin{quote} \bf}
{\end{quote}}

\title{The turbulent cascade in five dimensions} 
\author
{Jos\'e I. Cardesa$^{1\ast}$, Alberto Vela-Mart\'in$^{1}$, Javier Jim\'enez$^{1}$\\
\\
\normalsize{$^{1}$School of Aeronautics, Universidad Polit\'ecnica de Madrid, 28040 Madrid, Spain}\\
\\
\normalsize{$^\ast$To whom correspondence should be addressed; E-mail:  ji.cardesa@upm.es}
}

\date{}

\begin{document} 

% Double-space the manuscript.

% \baselineskip24pt
% \baselineskip24pt

% Make the title.

\maketitle 

% Place your abstract within the special {sciabstract} environment.

\begin{sciabstract}

To the naked eye, turbulent flows exhibit whirls of many different sizes. 
To each size, or scale, corresponds a fraction of the total energy 
resulting from a cascade in five dimensions: scale, time and 
three-dimensional space. Understanding this process is critical to 
strategies for modeling geophysical and industrial flows. By tracking 
the flow regions containing energy in different scales, we have 
detected the statistical predominance of a cross-scale link whereby 
fluid lumps of energy at scale $\Delta$ appear within lumps of scale 
$2\Delta$ and die within those of scale $\Delta/2$. Our approach 
uncovers the energy cascade in a simple water-like fluid, 
offering insights for turbulence models while paving the way 
for similar analyses in conducting fluids, quantum fluids and plasmas.

\end{sciabstract}

% In setting up this template for *Science* papers, we've used both
% the \section* command and the \paragraph* command for topical
% divisions.  Which you use will of course depend on the type of paper
% you're writing.  Review Articles tend to have displayed headings, for
% which \section* is more appropriate; Research Articles, when they have
% formal topical divisions at all, tend to signal them with bold text
% that runs into the paragraph, for which \paragraph* is the right
% choice.  Either way, use the asterisk (*) modifier, as shown, to
% suppress numbering.

\paragraph*{}
Perhaps no other area of physics research has borne the influence 
of a rhyming verse more than turbulence, where Richardson’s 
\textit{``Big whirls have little whirls that feed on their velocity, 
and little whirls have lesser whirls and so on to viscosity''} 
\cite{richardson2007weather} is embedded in the seminal theory 
of Kolmogorov, Onsager, von Weizsäcker and Heisenberg 
\cite{K41,onsager1949statistical,weizsacker1948spektrum,
heisenberg1948statistischen}. 
The last three physicists transcribed the phenomenology in 
terms of wave numbers, which were 
to become the predominant tool in theoretical studies of the 
energy cascade \cite{obukhoff1941energy,kraichnan1959structure,
leith1967diffusion,proudman1954decay,tatsumi1957theory}. 
Consequently, scale and wavenumber became 
almost interchangeable concepts. A crucial point in the development 
of theories was the scale locality of the cascade, understood in terms 
of how close wavenumbers are when energy is exchanged between them 
\cite{kraichnan1971inertial}. 
Since the advent of computer simulations, the locality of these wavenumber 
interactions has been controversial, with studies claiming evidence in 
favor of \cite{doma93} or against \cite{YeungBrass91} it. Rigorous 
explanations proposed for these discrepancies \cite{eyink2009localnessI,
aluie2009localnessII} advocate for the classic scale-local view 
of the cascade. The debate, however, has turned predominantly around the 
equivalence between wavenumber and scale, ruling out any possibility of 
attributing the ongoing cascade to specific whirls visible where the flow 
actually evolves: the real space. Furthermore, computer simulations of 
industrial and atmospheric flows are carried out on numerical grids 
representing physical space and rely heavily on the modeling of the 
interaction between the resolved (large) and subgrid (small) scales 
\cite{sagaut2006large}. 

\paragraph*{}
Studies of the interscale energy transfer based on real-space quantities 
share one of two limitations. They either focus 
on a subset of the source or sink terms responsible for the changes in 
energy at a point \cite{Wan_Meneveau,cardesa2015temporal}, or they make 
no use of time thus precluding any dynamical information or knowledge of 
causality \cite{aoyama_2005,piomelli_1991,Meneveau_cerutti_PoF1998,
davidson2005identifying,hill2002exact}. Often both 
limitations are combined. A noteworthy exception found a delay in the 
peak of the correlation between energy at two different scales when 
following the larger-scale flow \cite{MeneveauLund}, suggesting that 
eddy structures transfer their energy to smaller scales. In the wake 
of that study, we aimed to follow individual eddy structures. This 
has become possible with modern data-storage facilities where flow 
simulations are preserved in a movie-like manner. Such data sets 
have enabled the verification of phenomenological descriptions that 
eventually feed into dynamical models. 

\paragraph*{}
We analyzed data from a direct numerical simulation of turbulence in a 
triply periodic cube, obtained by solving the Navier-Stokes equations 
for an incompressible fluid by means of a deterministically forced and 
statistically steady pseudo-spectral code \cite{SM}. An important length 
scale in turbulent flows, $\eta$, is given by 
$\eta=\nu^{3/4}/\varepsilon^{1/4}$, where $\nu$ is the kinematic 
viscosity and $\varepsilon$ is the mean rate of kinetic energy 
dissipation. This small-scale length is associated with the tiniest 
whirls of turbulence. 
Our $\left( 2\pi\right)^{3}$ computational domain spanned 
$\left(1516\eta\right)^{3}$ in space and lasted $2090\tau$ small-scale 
time units $\tau=\sqrt{\nu/\varepsilon}$. Expressed in 
terms of large-scale length and time units $L_{int}$ and $T_{int}$ 
\cite{SM}, respectively, the simulation spanned $\left(5.3L_{int}\right)^{3}$
and $66T_{int}$, with snapshots stored every $0.078\tau$. Although previous
simulations surpassed our Reynolds number $L_{int}/\eta=284$, our long yet 
temporally resolved dataset with a sizeable scale separation allowed us 
to statistically characterize a phenomenon by tracking many flow regions 
throughout their life.

\begin{figure}[hb!]
  \begin{center}
    \includegraphics[trim = 0cm 0cm 0cm 0cm,width=13.25cm]{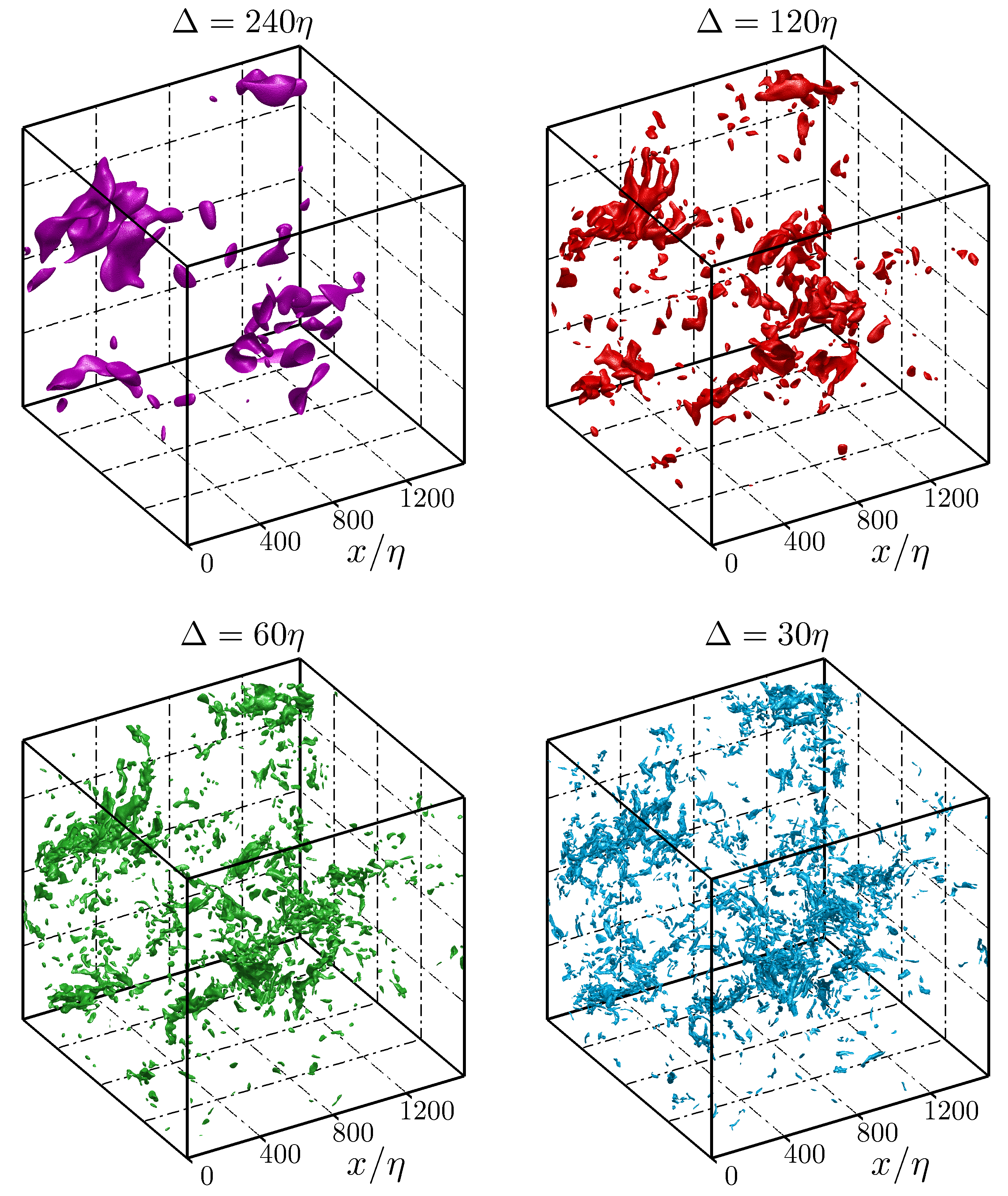}
  \end{center}
%  \noindent {\bf Fig. 1.}
  \caption{\textbf{Energy-eddies at four different scales $\Delta$ for the same instant 
  in a numerical simulation of turbulence in a periodic cube.} A 
  time sequence is shown in \href{https://torroja.dmt.upm.es/turbdata/Isotropic}{Movie S1} 
  \cite{SM}. The flow structures 
  observed are the spatially connected regions of the flow where the 
  energy at scale $\Delta$ is above a certain threshold \cite{SM}.}
  \label{fig:1} 
\end{figure}

\paragraph*{}
The tracked flow regions that we now introduce in detail underpin our 
definition of whirls, or eddies. 
We isolated a range of scales by 
filtering the velocity fields with a spatial band-pass filter. 
Owing to the homogeneity of the flow, we used an isotropic filter 
to simplify the concept of scale to a single scalar $\Delta$. 
We set the center of the filter band at the chosen scale $\Delta$
and used bands of constant width on a logarithmic scale 
\cite{aluie2009localnessII}. The upper and lower 
edges of the band resulted from subtracting two low-pass Gaussian 
filters \cite{SM}. We focused on four scales from the geometric sequence 
$\Delta/\eta = (30,60,120,240)$.  
This led to four time series of the dynamics of the flow, one for 
each scale. The object of our study was a scalar 
quantity, the kinetic energy, which evolved in time $t$, scale $\Delta$, 
and three-dimensional space $(x,y,z)$. The kinetic energy at a scale is 
the sum of the squared filtered velocity components. The flow structures in 
\Figref{fig:1} are geometrically connected regions of space 
where the energy is 
above a given threshold [\href{https://torroja.dmt.upm.es/turbdata/Isotropic}{Movie S1} 
\cite{SM}]. We chose the threshold 
systematically in the same way for all scales on the basis of the percolation 
properties of the energy at that scale \cite{SM,moisy2004geometry}. 
We further time-tracked these flow objects using a technique developed for the 
tracking of coherent structures in turbulent channel flows 
\cite{lozano2014time}. 
Whereas generally an object was born small as the underlying energy 
exceeded the threshold and died small as its intensity decreased, 
an object often merged with or split from other objects during its 
life. We grouped objects related by a connection at some point in 
space-time within the same temporal graph \cite{lozano2014time}. 
We called a graph emerging from this grouping an energy-eddy, 
or simply eddy, because it was educed according to the intense kinetic 
energy that it traced. 
We defined the scale of the eddy as the center $\Delta$ of the filter 
band that we used to compute it, the volume $(V)$ as the sum of volumes of 
all objects of a graph existing at a given instant, and the lifetime 
$(T_{life})$ as the time elapsed between the birth and death of the first 
and last object within the graph, respectively.

\begin{figure}
  \begin{center}
    \includegraphics[trim = 0cm 0cm 0cm 0cm,height=6.5cm]{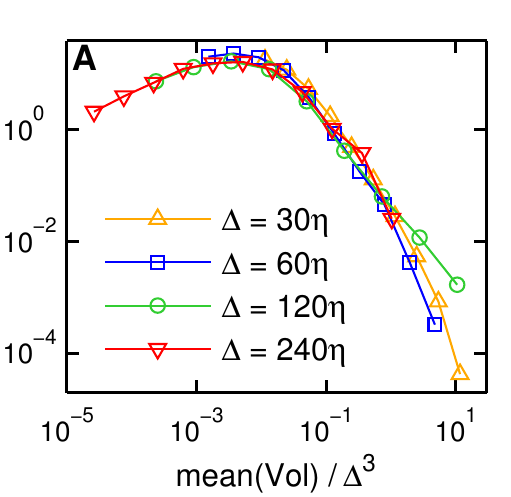} 
    \includegraphics[trim = -1.cm 0cm 0cm 0cm,height=6.275cm]{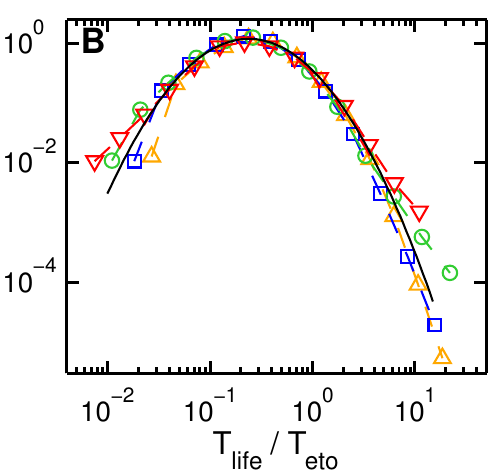} 
  \end{center}
%  \noindent {\bf Fig. 2.} 
  \caption{\textbf{Mean volumes and lifetimes of the eddies.}
    Probability density functions of \textbf{(A)} the mean volumes and 
    \textbf{(B)} 
    the lifetimes of the eddies. The solid black line in \textbf{(B)} follows 
    a log-normal distribution with a mean of $0.8$ and a standard deviation 
    of $1.3$. 
    The volumes use for \textbf{(A)} are the average over each eddy's 
    lifetime.}
  \label{fig:2} 
\end{figure}

\paragraph*{}
The volume distributions collapsed onto a single curve over a fairly wide 
range (\Figref{fig:2}A) with scaling parameter $\Delta$, after neglecting the tails. 
The lifetime distributions scaled with the local eddy turnover 
time $T_{eto}=\Delta^{2/3}\varepsilon^{-1/3}$ (\Figref{fig:2}B), 
found by assuming 
that for a range of scales where $L_{int}>>\Delta>>\eta$, 
the only relevant parameters are $\varepsilon$ and $\Delta$ 
\cite{K41}. The collapse of the 
probability density functions (\Figref{fig:2}B) supports that 
$T_{eto}$ is indeed a 
scaling parameter 
over a range of lifetimes, excepting the short-lived 
small-scale eddies for 
which viscosity cannot be neglected. A log-normal distribution resulted 
from a nonlinear fit to all our data (\Figref{fig:2}B), where the 
mean and 
standard deviation were the fitting parameters minimizing the difference. 
The parameters were $0.8$ and $1.3$, respectively, so that the average 
eddy lifetime was of the order of $T_{eto}$. Overall, the picture that 
emerges lends support to the eddy definition based on the 
observed scaling properties of the eddy sizes and lifetimes.

\begin{figure}
  \begin{center}
    \includegraphics[trim = 0cm 0cm 0cm 0cm,width=13.5cm]{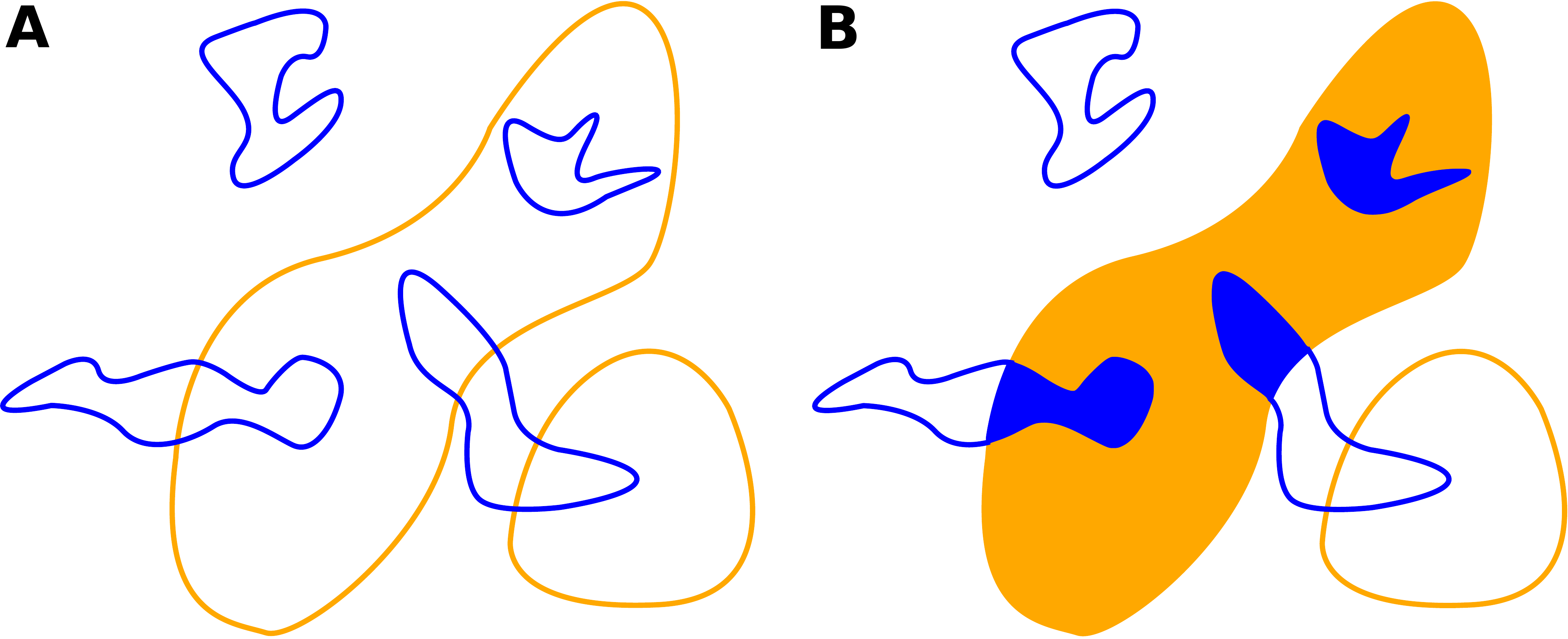}
  \end{center}
  % \noindent {\bf Fig. 3.} 
  \caption{\textbf{Sketch illustrating the intersection ratio between eddies of scale $A$ 
  (orange) and $B$ (blue)}. $R \left( A,B \right)$ for the largest of the 
  two eddies of scale $A$ in \textbf{(A)} is given by the sum of the 
  blue areas in \textbf{(B)}, divided by the union of the blue and orange 
  areas in \textbf{(B)}.}
  \label{fig:3} 
\end{figure}

\paragraph*{}
Our four time sequences provide the space-time position of the eddies 
at four scales. By superposing contemporary fields from any two scales 
at a time, we linked those two scales by computing their intersection 
in space. We defined the intersection ratio for a single eddy of 
scale $A$ intersecting $N$ eddies of scale $B$ at a given instant
\begin{equation}
R \left( A,B \right) \equiv 
\frac{\sum\limits_{i}^{N}V_{i}\left( A \cap B  \right)}{V\left( A \right)}, 
\end{equation}
where $V\left(A\right)$ is the volume of the eddy of scale $A$, 
and $V_{i}\left( A \cap B  \right)$ is the intersection 
volume between the eddy of scale $A$ and the $i$th intersected eddy of 
scale $B$ (\Figref{fig:3}). This ratio is unity if the eddy of scale $A$ is 
fully contained within one or more eddies of scale $B$, whereas it vanishes 
when $N=0$ in the case of no intersections. We thus quantified how the field 
of scale-$B$ eddies filled up individual eddies of scale $A$, which we 
followed.

\begin{figure}[t!]
  \begin{center}
    \includegraphics[trim = 0cm 0cm 0cm 0cm,height=6cm]{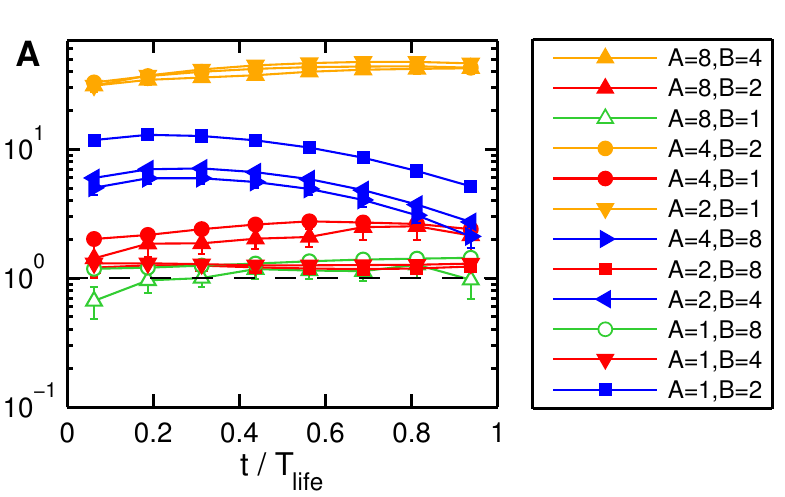} 
    \includegraphics[trim = 0cm 0cm 0cm 0cm,height=5.85cm]{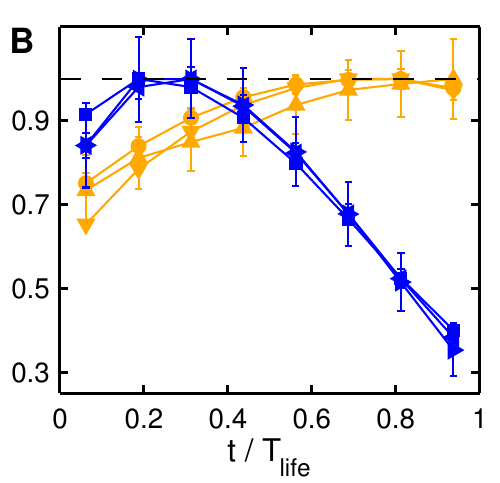}
  \end{center}
  % \noindent {\bf Fig. 4.}
  \caption{\textbf{Mean intersection ratio $R_{m}\left( A,B \right)$ between scales 
  $A$ and $B$ at different stages of the lifetime of scale-$A$ eddies.}
  Scales $\{1,2,4,8\}$ in the legend correspond to $\Delta/\eta = \{30,60,120,240\}$, 
  respectively. \textbf{(A)} Normalized by the intersection level obtained 
  by randomly locating fields $B$ with respect to $A$ 
  (null hypothesis) \cite{SM}. Unity implies a random intersection level.
  \textbf{(B)} Same as \textbf{(A)} but normalized by the maximum,
  keeping only $A$ and $B$ combinations separated by a factor of $2$. 
  Error bars in \textbf{(A)} and 
  \textbf{(B)} represent $95\%$ confidence intervals 
  \cite{benedict1996towards}. The horizontal dashed line is located at unity.}
  \label{fig:4} 
\end{figure}

\paragraph*{}
We considered whether the intersections depended on the scale and on the stage 
in the life of the eddies. We split the lifetime of each scale-$A$ eddy into 
equal fractions, or life stages. We then computed the mean intersection 
ratio $R_{m} \left( A,B \right)$ conditioned to a given life stage 
\cite{SM}. We normalized the result 
by the corresponding null hypothesis (\Figref{fig:4}A), 
which showed intersection 
levels higher than random for those scale combinations separated by a 
factor of $2$ (orange and blue curves). The intersections between eddies 
of scales further apart did not show a trend that was distinctly different from 
the null hypothesis, indicating little or no spatial overlap. We then 
normalized the scale combinations separated by a factor of $2$ with their 
corresponding maximum (\Figref{fig:4}B), 
which revealed that curves based on 
combinations where $A = 2B$ peaked towards the death of the scale-$A$ 
eddies, whereas with $A = B/2$, the peak was closer to their birth. This 
statistical signature shows unequivocally how the eddies of a given 
scale originate from eddies of twice their scale, whereas they give rise 
to eddies half their scale before disappearing. This process replicated 
successively through four scales separated by a factor of $2$, which 
suggests that a scale-local progression of the energy from the large 
to the small scales is, at the very least, a transited cascade path in 
homogeneous three-dimensional turbulence.

\paragraph*{}
The results show that energy is transferred to the smaller scales overall. 
This average trend is often labelled as a forward, or direct, cascade. 
Some individual eddies, however, follow opposite paths. To illustrate 
this point, we used a crude definition of a forward cascade event as an 
individual eddy of scale $A$ for which $R \left( A,B \right)$ averaged 
over the first half of $T_{life}$ is larger than $R \left( A,B \right)$ 
averaged over the second half of $T_{life}$, taking $A = B/2$. This 
occurred twice as often as backscatter events, defined as eddies 
for which $R \left( A,B \right)$ was larger during the second half of 
$T_{life}$, where $A = B/2$. 
Even though this way of quantifying the predominance of forward cascade 
versus backscatter is somewhat simplistic, it is important to 
keep such event counts in mind because the trends that we observed 
(\Figref{fig:4})
should not obscure the underlying bidirectionality. Large-eddy simulations, 
which are extensively used in engineering and meteorological contexts, 
require particular attention to the direction of the energy flux between 
the resolved and subgrid scales \cite{sagaut2006large}.

\paragraph*{}
Our analysis illustrated the locality of the
energy exchanges and, more broadly, the phenomenology of the 
turbulent cascade. We looked for and verified this 
statistically in physical space
with individual eddies of scales separated by factors of 
2, 4, and 8. Future work could include
studying the interaction between eddies separated
by a factor of 3 or 1.5, together with a simulation 
at larger scale separations where the
smallest scale is outside the viscous regime.
Moreover, our approach has led to an observation and 
not to the identification of the physical process causing it. 
We can now, however, design a study targeting those regions 
in space-time where the energy cascade takes place to 
understand the causes.
Lastly, our method was applied to a simple flow,
but nothing prevents its use in more complicated
cases with additional nonlinear terms.      
The presence of rotation, compressibility,
conductivity, and quantum effects
complicates the energy cascade considerably,
leading to areas of turbulence research where
Richardson's verse is a more limited part of the story.
Our approach could yield advances in
those fields, particularly as the required data
become available through numerical 
simulations \cite{perez2012energy,navon2016emergence,wan2016intermittency}.
% Our analysis illustrated the locality of the energy 
% exchanges and more broadly the phenomenology of the 
% turbulent cascade. We looked for and verified this 
% statistically in physical space with individual eddies 
% of scales separated by factors of $2$, $4$ and $8$. Future 
% work could include studying the interaction between 
% eddies separated by other factors ($3$ or $1.5$), together 
% with a simulation at larger scale separation where the 
% smallest scale is outside the viscous regime. Moreover, 
% since the approach we present is designed to observe the 
% presence of a statistical trend, we cannot infer what 
% physical process causes it. We can now, however, select 
% those parts of the flow where the energy transfer is 
% taking place and condition a more in-depth study of the 
% physical processes to those regions of space-time. 
% Finally, our method was applied to an incompressible fluid, 
% but nothing prevents its use in more complicated flows with 
% additional non-linear terms. The presence of rotation, 
% compressibility, conductivity and quantum effects 
% complicates the energy cascade considerably, leading to 
% communities of turbulence research where Richardson's 
% verse is an incomplete part of the story. We believe 
% our approach could lead to advances in those fields, 
% particularly as the required data becomes available 
% through numerical simulations 
% \cite{perez2012energy,navon2016emergence,wan2016intermittency}.

%\section*{Figures}

% %BibTeX users: After compilation, comment out the following two lines and paste in
% % the generated .bbl file. 

\clearpage
% \newpage

\bibliography{Cardesa.bbl}
\bibliographystyle{Science}

% \newpage
\section*{Acknowledgments}
We thank A. Lozano-Dur\'an for providing the code for the tracking of 
flow structures. All authors acknowledge funding from project COTURB 
(Coherent Structures in Wall-bounded Turbulence) of the 
European Research Council. The data for this study were 
obtained using the DECI-PRACE (Distributed European Computing 
Initiative-Partnership for Advanced Computing in Europe) 
resource Minotauro based in Spain at 
the Barcelona Supercomputing Centre. M.P. Encinar helped making the 
data public. The flow fields are freely 
available from our online database at
\url{https://torroja.dmt.upm.es/turbdata/Isotropic}.

%Here you should list the contents of your Supplementary Materials -- below is an example. 
%You should include a list of Supplementary figures, Tables, and any references that appear only in the SM. 
%Note that the reference numbering continues from the main text to the SM.
% In the example below, Refs. 4-10 were cited only in the SM.     
\section*{List of supplementary materials}
Materials and Methods\\
\Figrefs{figSM:1} to \ref{figSM:6}\\
\Tabref{tab:s1} \\
References \cite{OrszagPatterson,pope2000turbulent}\\
\href{https://torroja.dmt.upm.es/turbdata/Isotropic}{Movie 1}

\clearpage
\newpage

% \clearpage
% \newpage

\section*{Supplementary materials}

\subsection*{Materials and Methods}

\paragraph*{Numerical simulation}
The Navier-Stokes equations for an incompressible Newtonian fluid
are given by 
\begin{align}\nonumber
  \left( \frac{\partial}{\partial t} + u_{j}\frac{\partial}{\partial x_{j}}
  \right)u_{i} &= 
                 - \frac{\partial}{\partial x_{i}}(p/\rho) + 
                 \nu\frac{\partial^{2} u_{i}}{\partial x_{j}\partial x_{j}} + f_{i}\\
  \nonumber
  \frac{\partial u_{i}}{\partial x_{i}}&=0,
\end{align}
where $u_{i}$ and $f_{i}$ are the $i$-$th$ components of the velocity
and forcing, respectively, $p$ is the pressure,
$\rho$ is the density, 
$\nu$ is the kinematic viscosity,
$t$ is time, $x_{i}$ spans the three spatial directions and summation over
repeated indices is implied.
Our solver integrated a Fourier-transformed version of these equations 
using a third-order Runge-Kutta scheme
in time and a pseudo-spectral treatment of the  
triply periodic domain\cite{OrszagPatterson}. 
Each side span $L_{d}=2\pi$ and was
discretized with $N=1024$ collocation points. 
The forcing vector $\widehat{f}_i$, where the hat denotes Fourier
transform, followed
\begin{equation} \nonumber
  \widehat{f}_{i}\left(\boldsymbol{k},t\right)=
  \begin{cases} 
    \varepsilon \widehat{u}_{i}\left(\boldsymbol{k},t\right)/\left[2E_{f}(t)\right], 
    & \text{if $0 < k < k_{f}$},\\
    0, & \text{otherwise},
  \end{cases}
\end{equation}

where $\varepsilon\equiv1$ is the intended space-time mean 
dissipation $\frac{\nu}{2}
\left\langle 
\left( \frac{\partial u_{i}}{\partial x_{j}} 
+ \frac{\partial u_{j}}{\partial x_{i}} \right)^{2}
\right\rangle$,
$\boldsymbol{k}$ is the wave vector, $k=|\boldsymbol{k}|$,
$E_{f}(t) = \int_{0}^{k_{f}}E\left(k,t
\right) \dd k$ and $k_{f}=4\pi/L_{d}$. The spatial resolution
was set to $k_{max}\eta=2$ where $k_{max}\equiv\sqrt{2}N/3$, leading
to $Re_{\lambda} = 315$, $L_{int} = 1.18$. Quantities $E\left(k,t\right)$, 
$L_{int}$, and $Re_{\lambda}$ are defined as in equations
(6.188), (6.225) and (6.63) of Pope\cite{pope2000turbulent}. 
Quantity $T_{int}$ used in the
manuscript is defined as $T_{int}\equiv L_{int}/\sqrt{2K/3}$,
where $K\equiv\left\langle u_{i}u_{i}\right\rangle/2$ is the space-time 
mean kinetic energy. Phase-shifts were used for
the dealiasing. The numerical code was parallelized using a hybrid CUDA-MPI
implementation and ran on 64 NVIDIA GPUs in the Barcelona 
Supercomputing Centre (Spain). 

\begin{figure}[hb!]
  \begin{center}
    \includegraphics[trim = 0cmm 0cm 0cm 0cm,height=6.5cm]{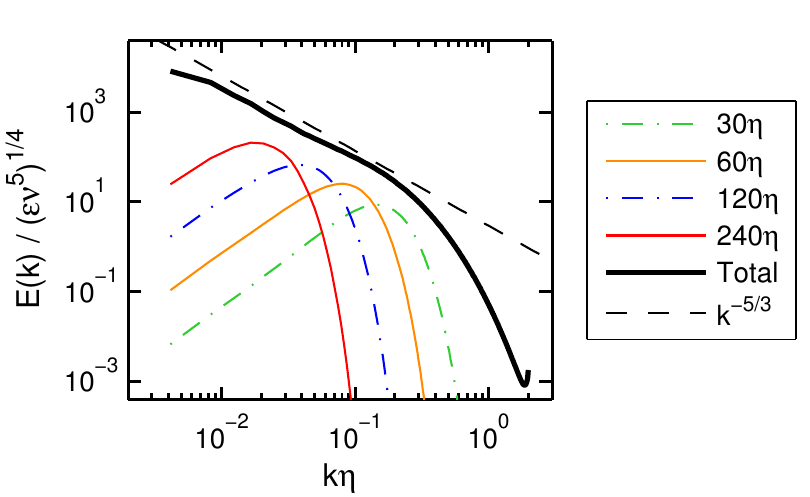}
  \end{center}
  \vspace{-0.5cm}
  % \noindent {\bf Fig. S1.} 
  \caption{Energy spectra of the band-pass filtered 
    velocities at different scales. Black thick solid line shows
    the energy spectrum of the unfiltered velocity.}\label{figSM:1}
\end{figure}

\paragraph*{Band-pass filtering}
To isolate a range of scales in the vicinity of $\Delta$, we combined 
two Gaussian low-pass filters with filter widths $\Delta_{1}$ and 
$\Delta_{2}$ as follows:
\begin{equation} \nonumber
  G_{_{\Delta}}\left( r \right) = \left(\frac{10}{\pi} \right)^{3/2}
  \left[ 
    \frac{1}{\Delta_{1}^{^{3}}}\exp\left(-\frac{10r^{^{2}}}{\Delta_{1}^{^{2}}}\right)  
    -
    \frac{1}{\Delta_{2}^{^{3}}}\exp\left(-\frac{10r^{^{2}}}{\Delta_{2}^{^{2}}}\right) 
  \right],
\end{equation}
where
\begin{equation} \nonumber
  \Delta_{_{1}} < \Delta < \Delta_{_{2}}~~;~~
  \Delta_{_{2}} =2\Delta_{_{1}}~~;~~
  \Delta=\sqrt{2}\Delta_{_{1}}~~;~~
  r^{2} = r_{i}r_{i}.
\end{equation}
The filtered velocity components 
$\overline{u}_{i}\left( \boldsymbol{x} \right) = \int G_{\Delta}\left( 
r \right)
u_{i}\left( \boldsymbol{x}-\boldsymbol{r}\right) d\boldsymbol{r}$ were
computed in Fourier space as 
$\overline{\widehat{u}}_{i}\left( \boldsymbol{k} \right) = \widehat{G}_{\Delta}\left( 
k \right)\widehat{u}_{i}\left( \boldsymbol{k} \right)$, where 
$\widehat{G}_{\Delta}\left( k \right) = 
\exp\left( -(k\Delta_{1})^{2}/40  \right)-
\exp\left( -(k\Delta_{2})^{2}/40  \right)$. The spectra of the 
energy at each of the four chosen $\Delta$ are shown on
\Figref{figSM:1}.

\begin{figure}[h!]
  \begin{center}
    \includegraphics[trim = 0cmm 0cm 0cm 0cm,height=6.5cm]{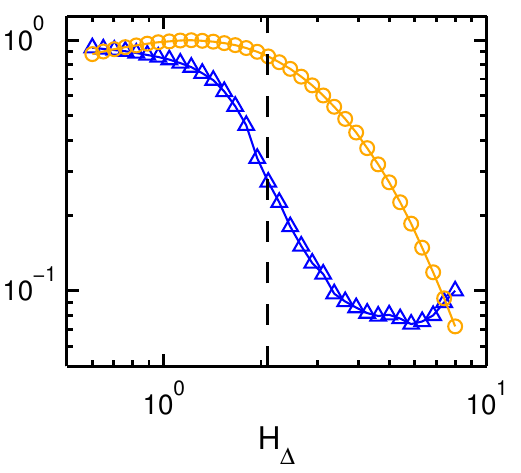} 
  \end{center}
  \vspace{-0.5cm}
  % \noindent {\bf Fig. S2.} 
  \caption{Percolation properties
    of the energy field at $\Delta = 60\eta$. Blue triangles show $\phi\left(H_{\Delta}\right)$
    and orangle circles show $\psi\left(H_{\Delta}\right)/max\left(\psi\right)$. 
    The vertical dashed line shows $\theta_{\Delta}^{*}$ for this $\Delta$.}
  \label{figSM:2}
\end{figure}

\begin{figure}[ht!]
  \begin{center}
    \includegraphics[trim = 0cm 0cm 0cm 0cm,height=6.0cm]{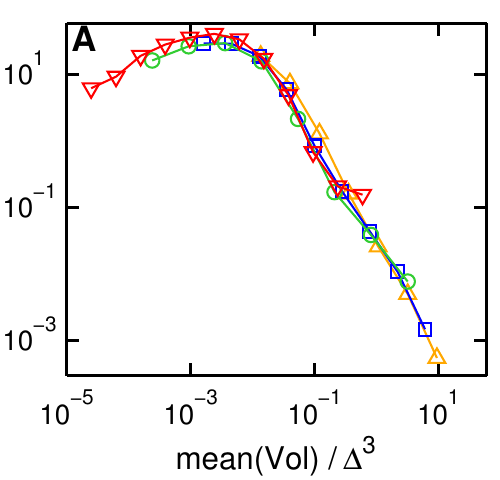}
    \includegraphics[trim = -2cm 0cm 0cm 0cm,height=6.0cm]{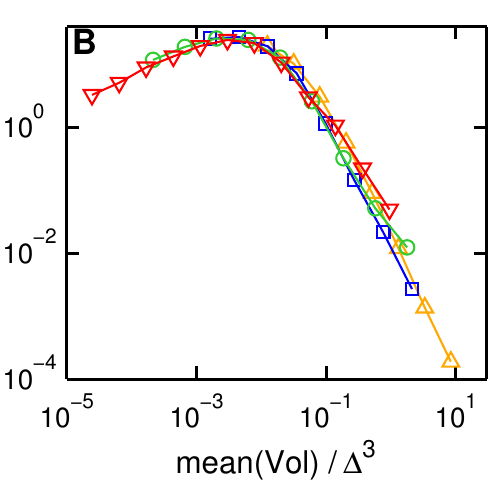}
  \end{center}
  \vspace{-0.7cm}
  % \noindent {\bf Fig. S3.} 
  \caption{Identical to \Figref{fig:2}A for thresholds 
    \textbf{(A)} $\theta_{\Delta}^{*}$ \textbf{(B)} $2^{1/2}\theta_{\Delta}^{*}$.}
  \label{figSM:3}
\end{figure}

\begin{figure}[h!]
  \begin{center}
    \includegraphics[trim = 0cm 0cm 0cm 0cm,height=6.0cm]{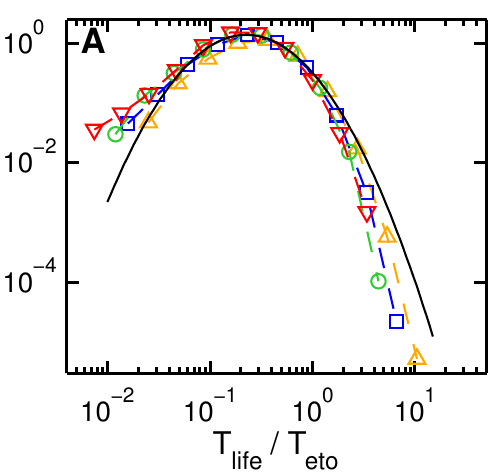}
    \includegraphics[trim = -2cm 0cm 0cm 0cm,height=6.0cm]{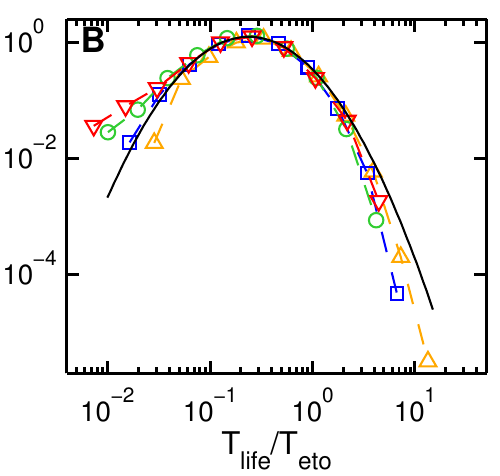}
  \end{center}
  \vspace{-0.7cm}
  % \noindent {\bf Fig. S4.} 
  \caption{Identical to \Figref{fig:2}B for thresholds 
    \textbf{(A)} $\theta_{\Delta}^{*}$ \textbf{(B)} $2^{1/2}\theta_{\Delta}^{*}$.}
  \label{figSM:4}
\end{figure}

\begin{figure}[h!]
  \begin{center}
    \includegraphics[trim = 0cm 0cm 0cm 0cm,height=6.0cm]{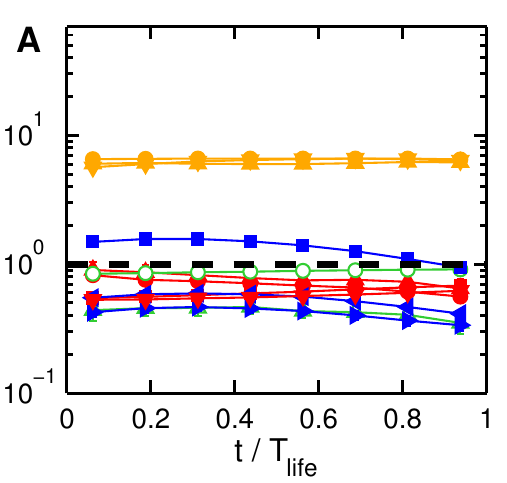}
    \includegraphics[trim = -2cm 0cm 0cm 0cm,height=6.0cm]{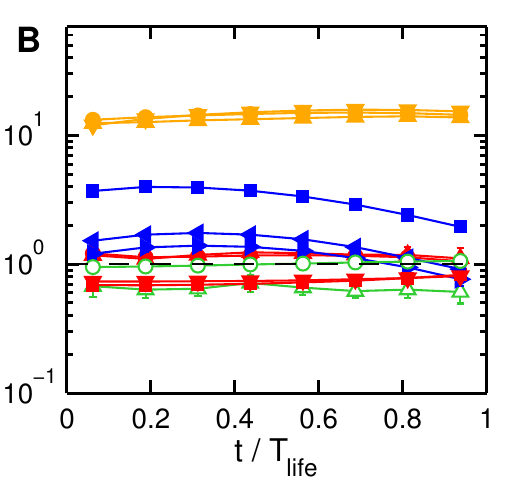}
  \end{center}
  \vspace{-0.7cm}
  % \noindent {\bf Fig. S5.} 
  \caption{Identical to \Figref{fig:4}A, with same color legend and for thresholds 
    \textbf{(A)} $\theta_{\Delta}^{*}$ \textbf{(B)} $2^{1/2}\theta_{\Delta}^{*}$.}
  \label{figSM:5}
\end{figure}

\begin{figure}
  \begin{center}
    \includegraphics[trim = 0cm 0cm 0cm 0cm,height=7.0cm]{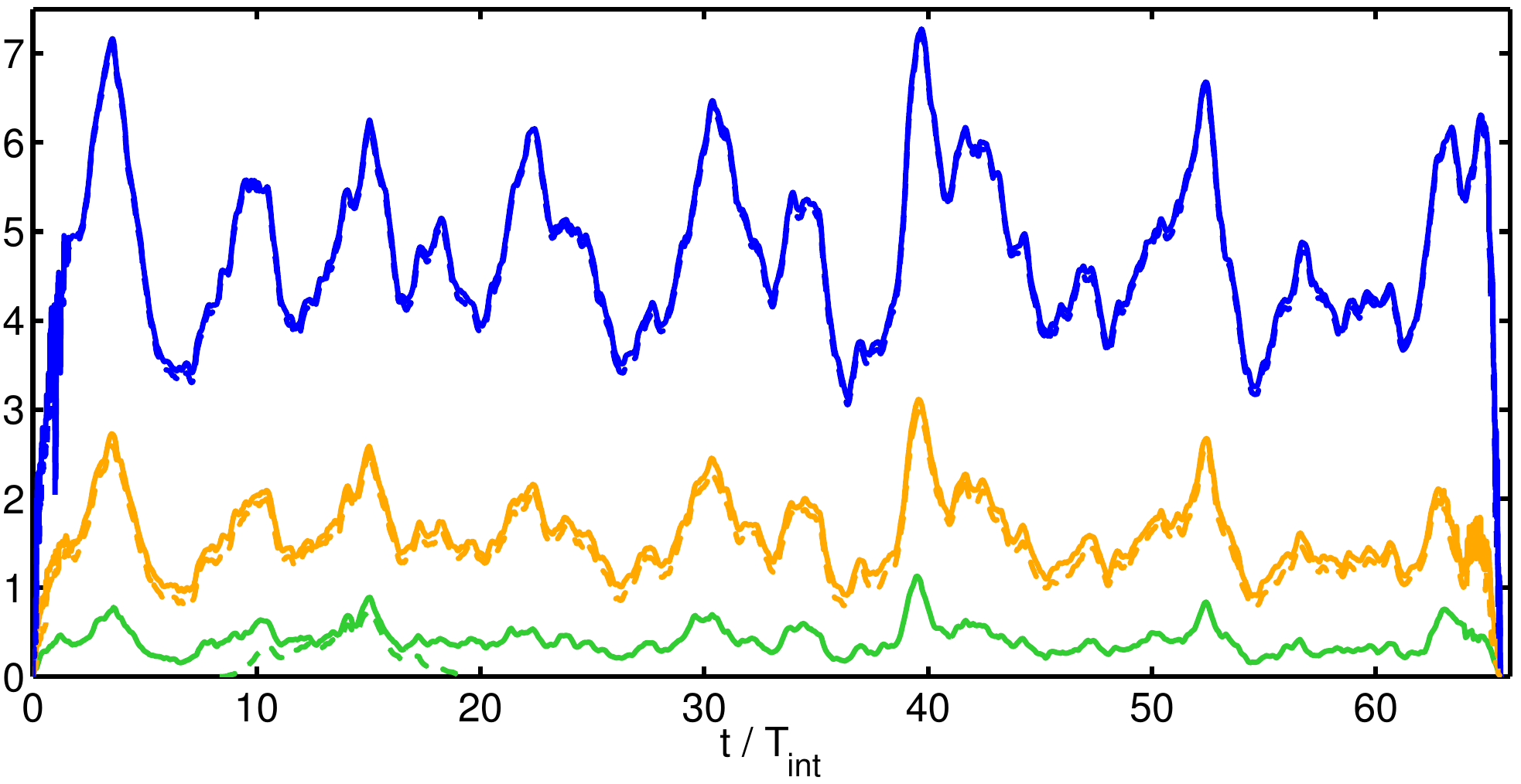}
  \end{center}
  \vspace{-0.5cm}
  % \noindent {\bf Fig. S6.} 
  \caption{Time-dependence of the volume of the largest graph (dotted lines) 
    and of the sum of all graphs (solid lines) during the simulation, 
    normalized by the domain volume $[\%]$. Different colors
    correspond to different thresholds:
    blue $\theta_{\Delta}^{*}$, orange ($2^{1/2}\theta_{\Delta}^{*}$),
    green ($2\theta_{\Delta}^{*}$). Results shown for scale
    $\Delta={120\eta}$, we found similar curves at the other $\Delta$.}
  \label{figSM:6}
\end{figure}

\paragraph*{Thresholding}
We studied the flow regions with energy above the 
threshold
$\theta_{\Delta} = \mu_{\Delta} + H_{\Delta}\sigma_{\Delta}$, 
where $\mu_{\Delta}$ and $\sigma_{\Delta}$ are the mean
and standard deviation, respectively, of the energy at 
scale $\Delta$. For a range of $H_{\Delta}$,
we gathered two quantities:
the ratio of the largest educed object divided by the sum 
of all objects, which we call $\phi$, and the number of 
educed objects $\psi$. The percolation
properties of the energy can be characterized\cite{moisy2004geometry} by 
$\phi\left(H_{\Delta}\right)$ and $\psi\left(H_{\Delta}\right)$, 
which we show on \Figref{figSM:2} for $\Delta/\eta=60$ 
- we found similar curves at the other $\Delta/\eta$. 
We define the critical $H_{\Delta}^{*}$ as the point where the 
rate of decrease in $\phi\left(H_{\Delta}\right)$ is greatest,
shown as a dashed vertical line on \Figref{figSM:2}.
The corresponding thresholds $\theta_{\Delta}^{*}$ were set as
reference, and we carried out the analysis reported in this 
manuscript at 3 thresholds for each scale:
$\theta_{\Delta} = \{\theta_{\Delta}^{*},\sqrt{2}\theta_{\Delta}^{*},
2\theta_{\Delta}^{*}\}$.
The educed structures contain a fraction of the flow's volume
and energy which are given in \Tabref{tab:s1}
for each $\theta_{\Delta}$ used, along with the corresponding number
of graphs. The results presented in the manuscript are all based 
on the highest threshold $2\theta_{\Delta}^{*}$. 
\Figrefstart{fig:2}A is reproduced at the two lower 
thresholds in \Figref{figSM:3}, while 
\Figref{fig:2}B is reproduced in 
\Figref{figSM:4} at the two lower thresholds. 
The conclusions drawn from these
two figures remain qualitatively unaltered by the threshold.
\Figrefstart{fig:4}A was found to depend on the threshold
as shown on \Figref{figSM:5}. We found the reason for this
dependency to be the creation, at the two lower thresholds,
of a single graph percolating the entire time series, with 
more than 90\% of the volume in the flow structures belonging
to the largest graph - see \Figref{figSM:6}. This influenced
significantly the intersection ratio computations due to most
structures being part of the same (largest) graph.

\paragraph*{Mean intersection $\boldsymbol{R_{m} \left( A,B \right)}$} 
We split $T_{life}$ for each scale-$A$ eddy 
into 8 equal fractions, or life stages, which are the same 
regardless of $T_{life}$.
We then compute for each scale-$A$ eddy the mean of $R \left( A,B \right)$ 
over each of its life stages,
obtaining a single 8-point series
$\left\langle R \left( A,B \right) \right\rangle$ for 
each eddy which corresponds to a set of life stages that is 
common across all scale-$A$ eddies.
In the next step, we compute the mean
of $\left\langle R \left( A,B \right) \right\rangle$ averaged
over all scale-$A$ eddies for a given life stage.
In this way we obtain the mean intersection ratio $R_{m} \left( A,B \right)$ 
as a function of where scale-$A$ eddies are with respect to their life.
We focus on those scale-$A$ eddies with lifetimes in the range 
$1/4<T_{life}/T_{eto}<2.5$,
which contains the PDFs on \Figref{fig:2}B from their mode to 
ten times the mode. 
This range keeps approximately 50\% of the graphs and the best collapsed 
portion of the PDFs.
We apply the lifetime restriction only to field $A$, 
while we keep all eddies from field $B$ when computing the intersections.
The trend in \Figref{fig:4}B persisted as we narrowed the range of 
admitted lifetimes for scale-$A$ eddies, so that comparing 
eddies of scale $A$ with lifetimes spread over a factor 
of up to 10 did not alter our main conclusion. 

\paragraph*{Null hypothesis of $\boldsymbol{R_{m} \left( A,B \right)}$}
The intersection ratio $R \left( A,B \right)$ 
between a scale-$A$ eddy
and a field of randomly located scale-$B$ eddies occupying $\sum V_{B}$ 
is taken as the volume fraction given by $\left(\sum V_{B}\right)/(8\pi^{3})$, where
$8\pi^{3}$ is the volume of the computational domain. 

\paragraph*{Temporal resolution} The temporal resolution between
the conserved velocity fields of the simulation was $0.078\tau$. 
We found, however, that one in every three fields could be used
for the tracking. With this coarser temporal resolution of 
$0.235\tau$,
less than $0.5\%$ of the total volume of the tracked structures
with the finest $\Delta$ was lost, while the
processing cost was significantly reduced.  

\newpage

\begin{table}
  \begin{center}
    \includegraphics[trim = 0cm 0cm 0cm 0cm,height=3.5cm]{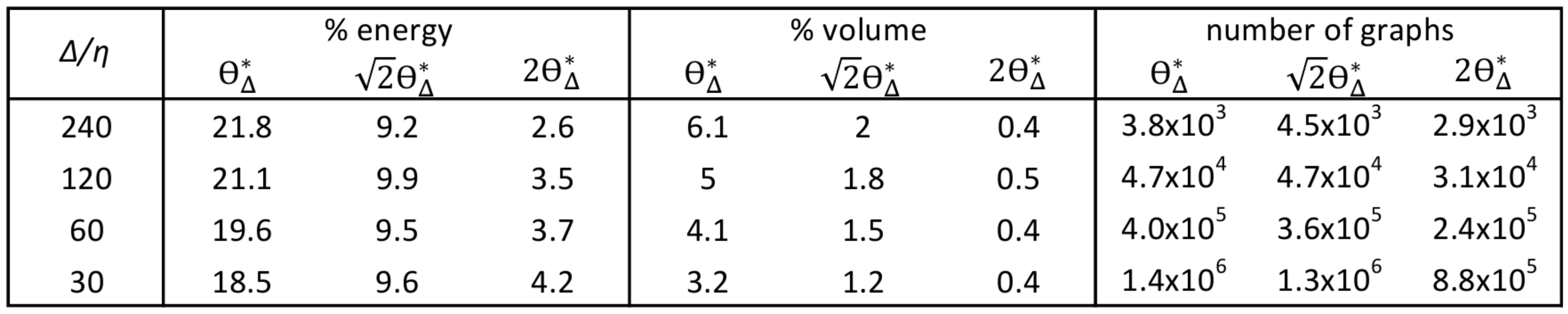} 
  \end{center}
  \vspace{-0.5cm}
  % \noindent {\bf Table S1.}
  \caption{Fraction of energy and volume contained in the educed objects 
  and number of educed graphs for the filter scales and thresholds used in our study.}
  \label{tab:s1}
\end{table}

\end{document}